\documentclass[aip,apl,reprint]{revtex4-1}
\usepackage{graphicx}

\begin{document}

\title{Substrate-induced bianisotropy in metamaterials}

\author{David A. Powell}
\email{david.a.powell@anu.edu.au}

\author{Yuri S. Kivshar}

\affiliation{Nonlinear Physics Centre, Research School of Physics and Engineering,\\
 Australian National University, Canberra ACT 0200, Australia }

\begin{abstract}
We demonstrate that the presence of a supporting substrate can break the symmetry of a metamaterial structure,
changing the symmetry of its effective parameters, and giving rise to bianisotropy.  This indicates that 
magneto-electric coupling will occur in all metamaterials fabricated on a substrate, including those with 
symmetric designs.
\end{abstract}

\maketitle

Optical metamaterials usually consist of a patterned metal-dielectric
composite structure mounted on a substrate. They are designed
to exhibit exotic properties, such as a negative refractive
index~\cite{Zhang2005}. The design process usually aims to achieve
the maximum possible symmetry, to
better approximate an ideal isotropic material~\cite{Baena2007}.
In fabricating optical metamaterials, it is important to have straight side walls,
since tapering breaks symmetry and causes bianisotropy.
A bianisotropic material is one which develops an electric polarization
in response to a magnetic field, and vice-versa, with material
parameters of the form $D_{i}=\epsilon_{0}\epsilon_{ij}E_{j}-i\xi_{ij}H_{j}/c$
and $B_{i}=\mu_{0}\mu_{ij}H_{j}+i\xi_{ij}^{T}E_{j}/c$. This
complicates the effective medium description, and
can inhibit the negative refractive index~\cite{Marques2002}.
The effect of fabrication imperfections has been studied in order
to characterise bianisotropy with the aim to reduce its effect~\cite{Ku2009a,Ku2009}.
However, in this Letter we show that \emph{the presence of a substrate intrinsically 
induces bianisotropy in a metamaterial}, which should be taken into account for
accurate characterization and control of the metamaterial properties.

A substrate has previously been shown to significantly
influence the plasmonic resonances associated with a negative refractive
index~\cite{Minovich2010b}. However, to date little attention has
been paid to the resultant changes in symmetry, an important exception
being {}``planar chiral'' structures \cite{Papakostas2003}. In
this case an essentially two-dimensional structure can exhibit optical
activity, either when the metamaterial array has low in-plane symmetry,
or when the sample is not normal to the incident wave-vector.
We will show here a different form of substrate-induced symmetry
breaking, for three-dimensional symmetric structures illuminated at normal incidence.

\begin{figure}
\includegraphics[width=1\columnwidth]{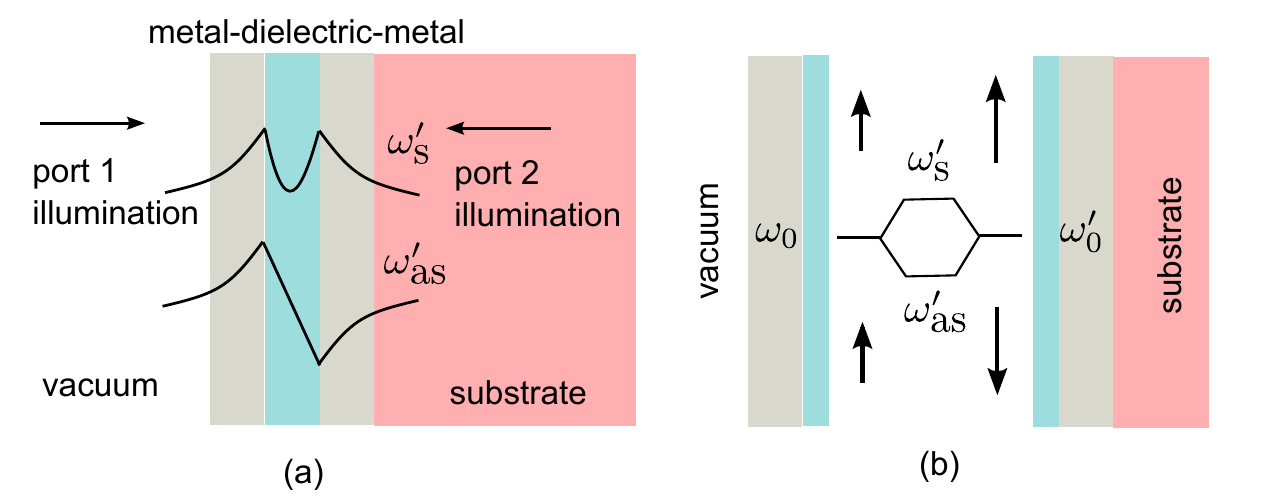}
\caption{(a) System under consideration, and (b) representation as hybridization
of non-identical resonators~\label{fig:config}}
\end{figure}

The negative index in optical metamaterials arises when there is both
a negative electric and magnetic response. The negative electric
response occurs naturally in metals, while the
negative magnetic response requires a pair of metal structures having an anti-symmetric
plasmonic mode. Thus we consider here a general model of a pair of
plasmonic resonators, which could be the two metal layers of a fishnet,
a cut-wire pair or a plasmonic dimer. In Fig.~\ref{fig:config}(a)
we show the structure under consideration, noting that the details
of this patterning are unimportant for this argument, except that
the pattern is uniform through the layers.

Conceptually, bringing the two identical patterned metallic layers
together will cause their resonances to hybridize into symmetric and
anti-symmetric modes. These will then respond to the imposed electric
and magnetic field respectively. In practice there is substantial
coupling of evanescent waves into the dielectric substrate, and this coupling
will be different for the two plasmonic layers.  Therefore we
can consider that the system is formed by the hybridization of two
non-identical plasmonic resonators, as shown in Fig.~\ref{fig:config}(b).
The nominally symmetric mode will have some anti-symmetric component,
and vice-versa. Each mode will therefore couple to both electric and
magnetic fields, resulting in bianisotropy. The argument is essentially
identical to that for intrinsically non-symmetric structures, once
it is appreciated that the substrate breaks the symmetry of the system,
even if it is outside the boundaries imposed during the retrieval
procedure. We will show a specific example of the experimentally important
fishnet structure. We consider the case of a semi-infinite substrate,
however the results still apply for a finite substrate.

The material parameters of a homogeneous structure can be retrieved
from transmission and reflection data\cite{Smith2002}. By utilizing the
two values of reflection from different sides of the structure, this
approach has been extended to asymmetric structures exhibiting bianisotropy
\cite{Li2009,Rill2008,Smith2005}. After calculating the generalized scattering parameters, we consider the
two values of the reflection coefficient
when illuminating from free space ($S_{11}$) or through the substrate
($S_{22}$). Even for a symmetric structure, these will not be equal,
since they are defined with respect to different values of the wave
impedance. However, we can transform the scattering matrix $S$ with
reference impedances $(\eta_{0},\eta_{sub})$ to $S'$ with both reference
impedances being $\eta_{0}$, and we find that for a simple dielectric
layer, the reflection coefficients become equal, and this should also
be true for a symmetric metamaterial.

\begin{figure}
\includegraphics[width=1\columnwidth]{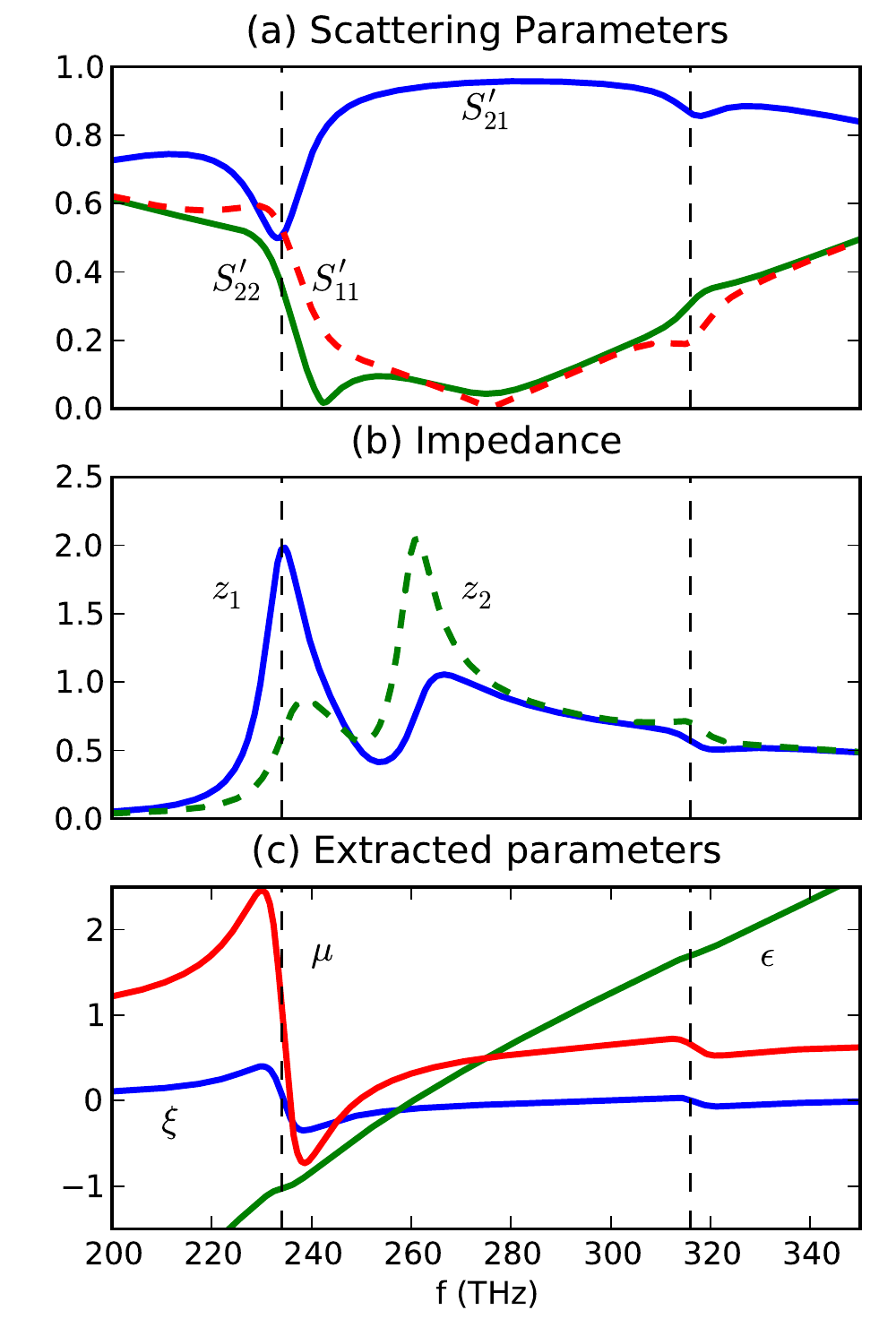}
\caption{Simulation results of the fishnet structure (a) Transmission and reflection
magnitude (b) real part of extracted impedance values (c) real part
of extracted permittivity, permeability, and magneto-electric coupling.
The dashed lines indicate the (approximately) anti-symmetric plasmonic
resonances. \label{fig:parameters}}
\end{figure}

We then apply this approach to a fishnet structure, with transverse
period of 400nm, consisting of 200$\times$350nm holes in metal-dielectric-metal
layers with metal thickness 30nm and dielectric thickness 30nm, on
a substrate of index 1.5 which is impedance matched to free space
(chosen for consistency with subsequent results, as justified below).
The structure was modeled in CST Microwave Studio \cite{CST}, with
the metal being modeled using a 4th order fit to the parameters of
silver, and the dielectric layer having $\epsilon=2.75$. The
transmission of this structure is shown in Fig.~\ref{fig:parameters}(a)
to exhibit two dips, corresponding to the excitation of approximately
anti-symmetric plasmonic modes, and a maximum corresponding to the
cut-off resonance of the hole mode \cite{Mary2008}. It can be clearly
seen that the two reflection coefficients are different in magnitude,
with the most significant differences corresponding to the frequencies
where plasmonic modes are excited. We
note that in modeling of lossless structures, bianisotropy may
not be immediately apparent, since $|S_{11}|=|S_{22}|$ is enforced
by energy conservation and reciprocity. However, the phase of these
two parameters may be altered by the substrate, hence bianisotropy
can still occur.

We apply the method given in Ref.~\onlinecite{Smith2005}
to extract the the two values of impedance, which are plotted in Fig.~\ref{fig:parameters}(b).
Failing to account for the bianisotropy in
the extraction results in some error in the effective index, and yields
an impedance value close to that of $z_{1}$ (not shown). It is clear
that the variation in impedance is strongest near the plasmonic
features, supporting the argument that the non-symmetric hybridization of
plasmonic modes is the cause of bianisotropy. In Fig.~\ref{fig:parameters}(c)
we extract the equivalent parameters of the structure using the method
from Ref.~\onlinecite{Li2009}, where for clarity only the real parts are shown.
It can be seen that the relevant component of the magneto-electric
coupling, $\xi_{xy}$ is significant around the magnetic resonance.
The observed values indicate that the magnetic polarization excited
by the electric field is about a third as strong as that excited by
the magnetic field. Interestingly, at $\sim$260THz,
which corresponds to the cut-off resonance of the holes where $\epsilon\approx0$
there is a strong variation between impedance values, but no corresponding
feature in the extracted magneto-electric coupling.

\begin{figure}
\includegraphics[width=0.8\columnwidth]{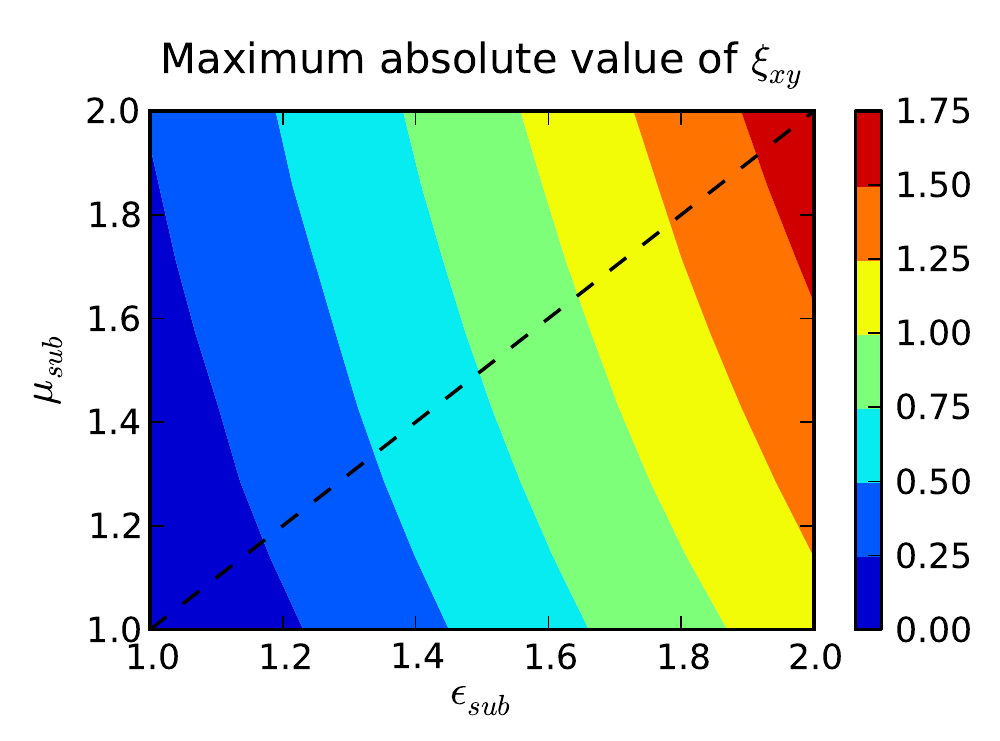}
\caption{Maximum real part of $\xi_{xy}$ as a function of substrate permittivity and permeability.
The dashed line indicates impedance matching to free space.\label{fig:swept_substrate}}
\end{figure}

In Fig.~\ref{fig:swept_substrate}
we plot the maximum magnitude of $\xi_{xy}$ over the considered
frequency range, as a function of the substrate parameters. Whilst in optical experiments the substrate
would have a purely dielectric response, we consider also a substrate
with a magnetic response to better understand the substrate influence. Increasing
$\epsilon_{sub}$ or $\mu_{sub}$ results in higher bianisotropy, although $\epsilon_{sub}$ has
a much greater influence.  The dashed line indicates the impedance match
between the substrate and free space, and the absence of any features in
this region indicates that bianisotropy
cannot be attributed to an impedance matching effect.  On the other hand,
for fixed impedance or permeability
bianisotropy increases monotonically with the index of the substrate.
This is consistent with our argument that the change in plasmon dispersion
is the cause of this effect.

While the extracted parameters reproduce the original transmission
and reflection data up to numerical accuracy, they are non-local in
nature \cite{Simovski2007a}. Therefore we shall probe the internal
field of a structure directly, in order to demonstrate that the magneto-electric
polarization is a real physical effect which does not rely on the
homogenizability of the structure. With reference to Fig.~\ref{fig:config}(a),
we excite the structure simultaneously from both directions with normally
incident waves. This creates a standing wave pattern, and by adjusting
the relative phase of the incident waves it is possible to position
the center of the structure at a zero of the electric or magnetic
field. Given that the thickness of the structure under consideration
is much less than the wavelength, we have a good approximation of
pure excitation by the magnetic or electric field respectively. In
order to simplify the excitation of the standing wave and the interpretation
of the results, we consider the case where the substrate impedance
is matched to free space, and the index is 1.5. We calculate the
induced magnetic and electric dipole moments by evaluating the integrals
$\bar{m}=j\omega/2\int\bar{x}\times(\epsilon-1)\bar{E}d^{3}x$ and
$\bar{p}=\int(\epsilon-1)\bar{E}d^{3}x$ over the metal and dielectric
regions.

\begin{figure}
\includegraphics[width=1\columnwidth]{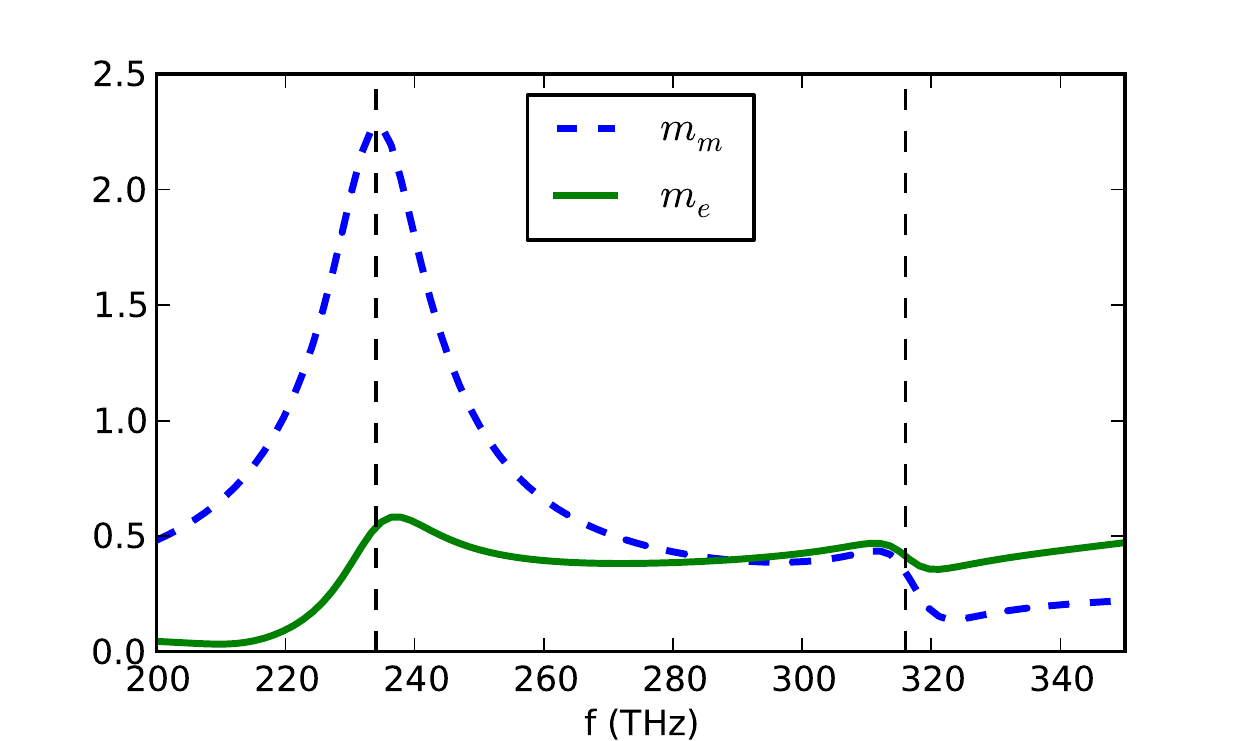}\caption{Magnitude of the magnetic
dipole moment induced in a fishnet under magnetic ($m_{m}$) and electric ($m_{e}$) excitation, with dashed
lines showing the plasmon resonance frequencies. \label{fig:polarisation}}
\end{figure}

The $x$ component of the induced magnetic dipole moment is shown
in Fig.~\ref{fig:polarisation} for excitation by an electric field
$m_{e}$, and is compared to $m_{m}$, the magnetic polarization induced
by the magnetic field. Quantities are normalized to the volume of
the unit cell, and to the incident magnetic field, thus they are dimensionally
equivalent to polarisabilities $\chi_{me}$ and $\chi_{mm}$.
It is clearly observable that
there are maxima of magneto-electric coupling corresponding to the
frequencies plasmon excitation observable in Fig.~\ref{fig:parameters}(a).
Significantly, at some frequencies, the magneto-electric polarization
is stronger than the purely magnetic polarization. In addition, there
is also some background cross-polarization which increases with frequency,
which is not observable in the effective parameters.

We expect that the effect predicted and analyzed here should occur in multi-layered
structures, however in this case the analysis becomes more complicated.
In particular, since this is an interface effect, the multi-valued
nature of the impedance should not strongly depend on the length of the structure. However,
inversion of the transmission and reflection assumes homogeneous parameters,
so the effect will be incorrectly attributed to bianisotropy throughout
the bulk. We note that approaches which account for
surface layers of a metamaterial exist~\cite{Simovski2007a}, but
are not widely utilized in the literature. We further propose that
at the interface between a bulk metamaterial and free-space,
the broken symmetry could also result in local bianisotropy. This
would not be observable by any technique based on S-parameter inversion of a finite thickness slab,
since the effect on both reflection coefficients would be identical.

In conclusion, we have demonstrated that the substrate breaks the
symmetry of structurally-symmetric metamaterials, resulting in significant 
bianisotropy. We have shown that this effect is strongest at the resonance of plasmonic modes,
which can be understood as the hybridization of non identical resonators to form the metamaterial.
Our results suggest that most metamaterials reported in the literature
will exhibit bianisotropy, and should be taken into account for
 accurate analysis of the metamaterial properties.

\begin{acknowledgments}
The authors thank Dr Ilya Shadrivov for fruitful discussions,
and also acknowledge support from the Australian Research Council.
\end{acknowledgments}


%

\end{document}